\documentclass[aip, pof, reprint]{revtex4-1}
\usepackage{graphicx}
\usepackage{color}
\usepackage{amssymb}
\usepackage{amsmath}
\usepackage{tabularx}
\usepackage{bm}
\usepackage{hyperref}
\usepackage{ltxtable}
\usepackage{natbib} 
\usepackage{longtable}
\usepackage{natbib}

\usepackage[usenames,dvipsnames]{xcolor}
\hypersetup{
colorlinks,
citecolor=blue,
filecolor=blue,
linkcolor=blue,
urlcolor=blue}

\newcommand{\be}{\begin{equation}}
\newcommand{\ee}{\end{equation}} 
\newcommand{\bea}{\begin{eqnarray}}
\newcommand{\eea}{\end{eqnarray}}

\begin{document}

\title{Similarities between the structure functions of thermal convection and hydrodynamic turbulence}

\author{Shashwat Bhattacharya}
\email{shabhatt@iitk.ac.in}
\affiliation{Department of Mechanical Engineering, Indian Institute of Technology Kanpur, Kanpur 208016, India}
\author{Shubhadeep Sadhukhan}
\affiliation{Department of Physics, Indian Institute of Technology Kanpur, Kanpur 208016, India}
\author{Anirban Guha}
\affiliation{School of Science and Engineering, University of Dundee, Dundee DD1 4HN, Scotland, United Kingdom}
\author{Mahendra K. Verma}
\affiliation{Department of Physics, Indian Institute of Technology Kanpur, Kanpur 208016, India}

\date{July 2019}%
\revised{October 2019}

\begin{abstract}
In this paper, we analyze the scaling of velocity structure functions of turbulent thermal convection. Using high-resolution numerical simulations, we show that the structure functions scale similar to those of hydrodynamic turbulence, with the scaling exponents in agreement with She and Leveque's predictions~[Phys. Rev. Lett. \textbf{72}, 336-339~(1994)].   The probability distribution functions of velocity increments are non-Gaussian with wide tails in the dissipative scales and become close to Gaussian in the inertial range. The tails of the probability distribution follow a stretched exponential.  We also show that in thermal convection, the energy flux in the inertial range is less than the viscous dissipation rate. This is unlike in hydrodynamic turbulence where the energy flux and the dissipation rate are equal.     
\end{abstract} 
\pacs{47.27.te, 47.27.-i, 47.55.P-}
\maketitle 

\section{Introduction}\label{sec:introduction}
Turbulence remains largely an unsolved problem for scientists and engineers even today. The energetics of three-dimensional homogeneous and isotropic turbulence is, however, well understood and was explained by \citet{Kolmogorov:DANS1941Dissipation,Kolmogorov:DANS1941Structure}. Here, the energy supplied at large scales cascades down to intermediate scales and then to dissipative scales. The rate of energy supply equals the energy flux, $\Pi_u$, and the viscous dissipation rate $\epsilon_u$. Kolmogorov showed that such flows exhibit the following property~\citep{Kolmogorov:DANS1941Dissipation,Kolmogorov:DANS1941Structure,Frisch:book}:
\bea
\langle [\{\mathbf{u(r+l)-u(r)}\}\cdot \mathbf{\hat{l}}]^3 \rangle &=& -\frac{4}{5}\Pi_u l, \nonumber \\
\Pi_u=\epsilon_u, \label{eq:Kolmogorov_4/5} 
\eea
for $\eta \ll l\ll L$, where $L$ is the length scale at which energy is supplied and is of the order of the domain size, and $\eta$ is the dissipative scale,  called Kolmogorov length scale. In Eq.~(\ref{eq:Kolmogorov_4/5}), $\langle . \rangle$ represents the ensemble average, and $\mathbf{u(r)}$ and $\mathbf{u(r+l)}$ are the velocity fields at positions $\mathbf{r}$ and $\mathbf{r+l}$ respectively. The left-hand side of Eq.~(\ref{eq:Kolmogorov_4/5}), denoted as $S_3^u(l)$, is the third-order velocity structure function. For any order $q$, one expects, using dimensional analysis, that $S_q^u(l)= \langle [\{\mathbf{u(r+l)-u(r)}\} \cdot \mathbf{\hat{l}}]^q \rangle \sim l^{q/3}$. Using the theory of \citet{Obukhov:1949} and \citet{Corrsin:JAP1951} on turbulence with passive scalar $\theta$, dimensional analysis yields $S_q^\theta(l) \sim l^{q/3}$, where $S_q^\theta(l) = \langle \{\theta(\mathbf{r+l}) - \theta(\mathbf{r})\}^q \rangle$ is the  structure function for the passive scalar. The aforementioned relations for $S_q^u$ and $S_q^\theta$ are known as Kolmogorov-Obukhov (KO) scaling in literature. In reality, however, the exponents deviate from $q/3$ (other than for 3) due to intermittency effects. The velocity structure functions scale as $S_q^u(l) \sim l^{\zeta_q}$, \textcolor{black}{where the exponents $\zeta_q$ fit well with the model of \citet{She:PRL1994}.}    

The scaling of structure functions of turbulent convection, however, remains an unsolved problem and hence is the theme of our present paper. We focus on Rayleigh-B\'{e}nard Convection (RBC) that deals with a fluid enclosed between two horizontal plates, with the bottom plate kept at a higher temperature than  the top plate. In thermal convection, complications arise due to anisotropy introduced by gravity, and also because the temperature $T$ is an active scalar. 

For stably stratified turbulence, \citet{Bolgiano:JGR1959} and \citet{Obukhov:DANS1959} predicted the kinetic energy spectrum $E_u(k$) and thermal energy spectrum $E_T(k)$ to scale as $k^{-11/5}$ and $k^{-7/5}$ respectively, where $k \sim 1/l$ is the wavenumber. An extension of Bolgiano-Obukov (BO) theory to structure functions gives $S_q^u(l) \sim l^{3q/5}$ and $S_q^T(l) \sim l^{q/5}$, where $S_q^T$ is the temperature structure function. BO scaling occurs above the Bolgiano length scale $l_B$, where the buoyancy forces are dominant. \textcolor{black}{Evidences of BO scaling have been observed in recent studies of stably-stratified~\cite{Kumar:PRE2014,Verma:NJP2017} and rotating stratified turbulence~\cite{Rosenberg:PF2015}}. Using theoretical arguments, \citet{Procaccia:PRL1989}, \citet{Lvov:PRL1991}, \citet{Lvov:PD1992}, and \citet{Rubinstein:NASA1994} proposed the applicability of BO scaling to RBC as well. Researchers have attempted to confirm the above theory with the help of experiments and numerical simulations, as well as using theoretical arguments.

\citet{Benzi:EPL1994,Benzi:EPL1994b} simulated both 2D and 3D RBC using Lattice Boltzmann method and computed velocity and temperature structure functions up to the sixth order. They could not observe any discernible scaling for the structure functions due to short inertial range. They found them, however, to be self-similar for a wide range of $l$, a phenomenon known as extended self-similarity (ESS)~\cite{Benzi:PRE1993,Chakraborty:JFM2010}. Further, they claimed BO scaling from the relationship between the velocity and the temperature structure functions. \citet{Ching:PRE2000} computed temperature and velocity structure functions of thermal convection using the experimental data of \citet*{Heslot:PRA1987}, and \citet*{Sano:PRA1989}, as well as the numerical data of \citet{Benzi:PD1996}. Although \citet{Ching:PRE2000} observed two distinct scaling regimes separated by the Bolgiano scale, the scaling exponents deviated from BO theory. 

Many researchers obtained KO scaling in the bulk and attributed it to the large value of local $l_B$, which is of the same order as the box size. Since $l_B$ is small near the walls, it is argued that the structure functions in those regions follow BO scaling. Using third-order structure functions calculated using their lattice Boltzmann simulation data, \citet{Calzavarini:PRE2002} claimed BO scaling near the walls and KO scaling at the cell center.  High-resolution multipoint measurements of velocity and temperature fields in water were conducted by \citet*{Sun:PRL2006}. Their exponents of velocity structure functions computed at the cell center fit well with the She-Leveque model, with the lower orders following Kolmogorov scaling. Using refined similarity hypothesis, \citet{Ching:PRE2013} derived power-law relations for conditional velocity and temperature structure functions computed at given values of the locally averaged thermal dissipation rate. \citet{Ching:PRE2013} further computed the conditional temperature structure functions up to the fourth order using the experimental data of \citet{He:PRE2009}. Based on the observed power-law scaling, they concluded BO scaling near walls and KO scaling at the cell center.

Using the experimental data of \citet{Castaing:JFM1989} and \citet{Shang:PRL2003}, \citet{Ching:PRE2007} computed the structure functions of plume velocity and found them to scale similar to the temperature structure functions. This is unlike the case of velocity structure functions in BO scaling, where they scale differently from the temperature structure functions. \citet{Kunnen:PRE2008} conducted direct numerical simulations of RBC in a grid resolution of $129 \times 257 \times 257$. The velocity structure functions computed by them follow BO scaling for Rayleigh number $\mathrm{Ra}=10^8$ and Kolmogorov scaling for higher $\mathrm{Ra}$. \citet{Ching:PRE2008Anomalous} calculated temperature structure functions using shell model of homogeneous RBC and found them to deviate significantly from BO scaling for $q>4$.  \citet{Kaczorowski:JFM2013} conducted direct numerical simulations (DNS) of RBC in grids ranging from $64^3$ to $770^3$, and found that the velocity structure functions computed at cell center approach Kolmogorov scaling for lower orders. 

From the conflicting nature of past results, it is clear that the behaviour of the structure functions of turbulent convection has not yet been clearly established. \citet{Lohse:ARFM2010} reviewed the experimental, numerical and theoretical results of past works critically and raised doubts on the applicability of BO scaling in RBC. Recently, using phenomenological arguments and numerical simulations, \citet{Kumar:PRE2014} and \citet{Verma:NJP2017} showed Kolmogorov energy spectrum in RBC. Using energetics arguments, they derived that the energy cascade rate in turbulent convection is constant, leading to Kolmogorov scaling. \textcolor{black}{Their predictions are being accepted and acknowledged by several groups as is evident from recent literature~\cite{Meuel:SR2018,Alexakis:PR2018,Bruneau:PRF2018,Shestakov:JAMTP2017,Pawar:PF2016,Bhattacharjee:PLA2015,Schumacher:PNAS2015}.  However, some researchers still believe that BO scaling is applicable to RBC~\cite{Rosenberg:PF2015,Shrestha:PRE2016,Rincon:AA2017,Rincon:LRSP2018}.} In this paper, using numerical simulations, we reinforce the results of \citet{Kumar:PRE2014} and \citet{Verma:NJP2017} by showing that the velocity structure functions of thermal convection scale similarly as those of 3D hydrodynamic turbulence. We further show that although the energy flux in turbulent convection is constant similar to hydrodynamic turbulence, it differs from viscous dissipation rate. We will discuss the scaling of temperature structure functions in a future work.

The outline of the paper is as follows: In Sec.~\ref{sec:Equations}, we describe the governing equations of RBC. In Sec.~\ref{sec:Phenomenology}, we discuss the phenomenology of turbulent convection and derive the scaling of third-order structure functions. In Sec.~\ref{sec:Numerics}, we briefly discuss the simulation details and the procedure employed to calculate the velocity structure functions. In Sec.~\ref{sec:SFu}, we present the scaling of the structure functions and discuss the nature of the probability distribution functions of velocity increments. Further, we compare the energy flux and viscous dissipation rate in RBC and show that the flux is less than the dissipation rate.  Finally, we conclude in Sec.~\ref{sec:Conclusions}. 

\section{Governing equations} \label{sec:Equations}
In RBC, under the Boussinesq approximation~\cite{Chandrasekhar:book:Instability,Chilla:EPJE2012}, we assume the kinematic viscosity $\nu$, thermal diffusivity $\kappa$, and thermal expansion coefficient $\alpha$ to be constants. Further, the density of the fluid is taken to be constant except for the buoyancy term in the momentum equation. The temperature field $T$ can be split as
\begin{equation}
T(x,y,z) = T_c(z) + \theta(x,y,z), \label{eq:Temp_Decomposition}
\end{equation}
where $T_c(z)$ is the conduction temperature profile, and $\theta(x,y,z)$ is the deviation of temperature from the conduction state. \textcolor{black}{Further, the temperature fluctuation $\theta$ is related to the density fluctuation $\rho$ as~\cite{Chandrasekhar:book:Instability,Verma:book:BDF}} 
\begin{equation}
\textcolor{black}{\rho = -\rho_0 \alpha \theta,} \nonumber
\end{equation} 
\textcolor{black}{where $\rho_0$ is the mean fluid density.}
The governing equations of RBC are as follows:
\bea
\frac{\partial \mathbf{u}}{\partial t} + (\mathbf{u} \cdot \nabla) \mathbf{u} & = & -\frac{\nabla \sigma}{\rho_0} + \alpha g \theta \hat{z} + \nu \nabla^2 \mathbf{u}, \label{eq:momentum}\\
\frac{\partial \theta}{\partial t} + (\mathbf{u} \cdot \nabla) \theta & = & \frac{\Delta}{d}u_z + \kappa \nabla^2 \theta, \label{eq:theta}  \\
\nabla \cdot \mathbf{u} & = & 0, \label{eq:continuity} 
\eea  
where $\mathbf{u}$ and $\sigma$ are the velocity and the pressure fields respectively, and $\Delta$ and $d$ are the temperature difference and distance respectively between the top and the bottom plates. 

Using $d$ as the length scale, $\sqrt{\alpha g \Delta d}$ as the velocity scale, and $\Delta$ as the temperature scale, we non-dimensionalize Eqs.~(\ref{eq:momentum})-(\ref{eq:continuity}), which yields
\bea
\frac{\partial \mathbf{u}}{\partial t} + \mathbf{u} \cdot \nabla \mathbf{u} &=& -\nabla \sigma + \theta \hat{z} +  \sqrt{\mathrm{\frac{Pr}{Ra}}}\nabla^2 \mathbf{u}, \label{eq:NDMomentum} \\
\frac{\partial \theta}{\partial t} + \mathbf{u}\cdot \nabla \theta &=& u_z + \frac{1}{\sqrt{\mathrm{Ra} \mathrm{Pr}}}\nabla^2 \theta, \label{eq:NDTheta}\\
\nabla \cdot \mathbf{u} &=& 0, \label{eq:NDContinuity}
\eea 
where $\mathrm{Ra}=\alpha g \Delta d^3/(\nu \kappa)$ is the Rayleigh number, and $\mathrm{Pr}=\nu/\kappa$ is the Prandtl number. The Rayleigh and Prandtl numbers are the main governing parameters of RBC. 

In the next section we construct a phenomenology for  the structure functions of turbulent convection.

\section{Hydrodynamic turbulence-like phenomenology for turbulent convection} \label{sec:Phenomenology}
\subsection{Energy fluxes and spectra in \textcolor{black}{hydrodynamic turbulence and thermal convection}} \label{subsec:Flux_spectrum}
For 3D hydrodynamic turbulence, the energy cascade rate $\Pi_u$ in turbulent flows is constant in the inertial range ($\eta \ll l \ll L$). Dimensional analysis gives the following relation for the energy spectrum $E_u(k)$: 
\begin{equation}
E_u(k)= K_\mathrm{KO} (\Pi_u)^{2/3}k^{-5/3},
\label{eq:KolmogorovSpectrum}
\end{equation}
where $K_{\mathrm{KO}}$ is the Kolmogorov constant. The aforementioned $k^{-5/3}$ spectrum is known as Kolmogorov's spectrum.
In this section, we briefly describe the phenomenological arguments of \citet{Kumar:PRE2014}, \citet{Verma:NJP2017}, and \citet{Verma:book:BDF}, according to which the energy spectrum in turbulent convection follows Kolmogorov scaling with constant energy flux, contrary to the arguments of \citet{Lvov:PRL1991} and \citet{Lvov:PD1992}, who propose Bolgiano-Obukhov scaling with variable flux.  

In all turbulent flows, the following can be derived using Eq.~(\ref{eq:momentum})~(see Refs.~\cite{Frisch:book,Lesieur:book:Turbulence,Verma:book:BDF}):
\begin{equation}
\frac{\partial}{\partial t}E_u(k,t)=-\frac{\partial}{\partial k}\Pi_u(k,t)+\hat{\mathcal{F}}(k,t)-\hat{D}_u(k,t),
\label{eq:energyBalance}
\end{equation}
where $\hat{\mathcal{F}}(k,t)$ is the energy feed due to forcing, and $\hat{D}_u(k)$ is the dissipation rate of kinetic energy. For a steady state, we have $\frac{\partial}{\partial t}E_u(k,t) \approx 0$ that modifies Eq.~(\ref{eq:energyBalance}) to
\begin{equation}
\frac{d}{dk}\Pi_u(k) = \hat{\mathcal{F}}(k)-\hat{D}_u(k).
\label{eq:energyBalance_Steady}
\end{equation}
Now, we will separately consider hydrodynamic turbulence and RBC and show that the flux is constant for both the cases. However, there is a difference between the two fluxes, as shown below.

\subsubsection{Hydrodynamic turbulence} \label{subsubsec:HydroPhenomenology}
The forcing in hydrodynamic turbulence is supplied at small wavenumbers. In the inertial range, $\hat{\mathcal{F}}(k)=0$ and $\hat{D}_u(k)$ is negligible. This results in the following~\citep{Kolmogorov:DANS1941Dissipation,Kolmogorov:DANS1941Structure,Lesieur:book:Turbulence,Frisch:book}:
\begin{equation} 
\frac{d}{dk} \Pi_u(k) = 0, \quad \Rightarrow \Pi_u(k) = \mathrm{constant}.
\label{eq:constFluxHydro} 
\end{equation}  
\textcolor{black}{Note that in hydrodynamic turbulence, the forcing injection $\mathcal{F}(k)$ is  modelled numerically in many ways.  Refer to \citet{Canuto:book:SpectralFluid} for details. } 

Let us consider a small wavenumber $k_0$ that lies in the inertial range and is slightly larger than the forcing wavenumber. \textcolor{black}{Integration of Eq.~(\ref{eq:energyBalance_Steady}) from $0$ to $k_0$ yields
\begin{equation}
\Pi_u(k_0) - \Pi_u(0) =   \int_{0}^{k_0} \hat{\mathcal{F}}(k)dk - \int_{0}^{k_0}\hat{D}_u(k)dk. \label{eq:Energy_integrate_0} \\
\end{equation}
Note that $\int_{0}^{k_0} \hat{\mathcal{F}}(k)dk$ is the total energy injection rate for hydrodynamic turbulence. Since $\Pi_u(0)=0$ and the dissipation at small wavenumbers is negligible, we obtain
\begin{equation}
\Pi_u(k_0)  \approx  \int_{0}^{k_0} \hat{\mathcal{F}}(k)dk. \label{eq:Flux_forcing_HT} \\
\end{equation}
Now,} integration of Eq.~(\ref{eq:energyBalance_Steady}) from $k_0$ to $\infty$ yields
\begin{equation}
\Pi_u(\infty) - \Pi_u(k_0) =   \int_{k_0}^{\infty} \hat{\mathcal{F}}(k)dk - \int_{k_0}^{\infty}\hat{D}_u(k)dk. \label{eq:Energy_integrate} \\
\end{equation}
Since $\Pi_u(\infty)=0$ and $\hat{\mathcal{F}}(k)=0$ for $k \in [k_0, \infty)$, we get
\begin{equation}
\Pi_u(k_0) = \int_{k_0}^{\infty}\hat{D}_u(k)dk \approx \int_{0}^{\infty}\hat{D}_u(k)dk = \epsilon_u.
\label{eq:Diss_flux_HT}
\end{equation} 
Note that $k_0$ is small, and $\hat{D}_u(k)$ is small in the forcing band.  Therefore the lower limit of the aforementioned integration has been replaced with 0. \textcolor{black}{Thus, using Eqs.~(\ref{eq:Flux_forcing_HT},\ref{eq:Diss_flux_HT}) we deduce that in hydrodynamic turbulence, the energy flux in the inertial range is constant, and is approximately equal to the dissipation rate $\epsilon_u$ and the total energy injection rate.} 

\begin{figure*}
\centering
\includegraphics[scale=0.2]{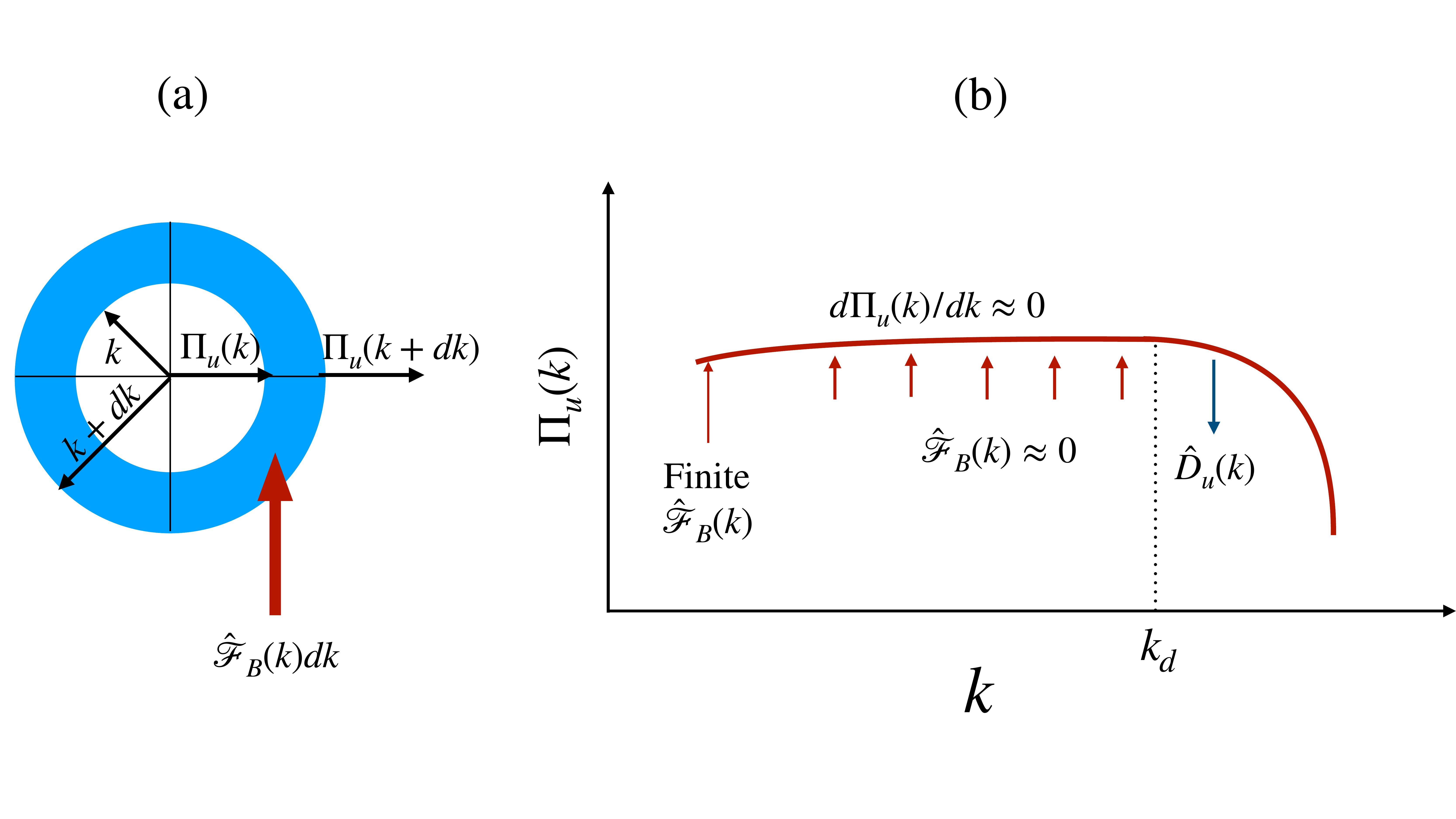}
\caption{For RBC: (a) A schematic diagram of a wavenumber shell of radius $k$ showing the buoyant energy feed $\hat{\mathcal{F}}_B$ and the kinetic energy flux $\Pi_u(k)$. (b) Schematic plot of $\Pi_u(k)$ vs. $k$. $\Pi_u(k) \approx \mathrm{constant}$ in the inertial range because of weak $\hat{\mathcal{F}}_B$. Viscous dissipation $\hat{D}_u(k)$ is dominant for $k>k_d$.}
\label{fig:Flux_RBC}
\end{figure*}

\subsubsection{Thermal convection} \label{subsubsec:ConvectionPhenomenology}
\textcolor{black}{In turbulent convection, the energy is injected into the system by buoyancy. We denote this energy feed as $\hat{\mathcal{F}}_B(k)$. Note that we do not inject energy externally in convection as we do in hydrodynamic turbulence. Further, unlike hydrodynamic turbulence, $\hat{\mathcal{F}}_B(k)$ acts at all scales in thermal convection. Replacing $\hat{\mathcal{F}}(k)$ with $\hat{\mathcal{F}}_B(k)$, we rewrite Eq.~(\ref{eq:energyBalance_Steady}) as
\begin{equation}
\frac{d}{dk}\Pi_u(k) = \hat{\mathcal{F}}_B(k)-\hat{D}_u(k).
\label{eq:energyBalance_Steady_RBC}
\end{equation}
} Since hot plumes ascend and the cold plumes descend, $u_z$ and $\theta$ are positively correlated, which means that~\citep{Kumar:PRE2014,Verma:NJP2017}
\begin{equation}
\langle \theta(\mathbf{r}) u_z(\mathbf{r}) \rangle > 0 \nonumber.
\end{equation}
Using this condition, \citet{Kumar:PRE2014} and \citet{Verma:NJP2017} claimed that $\hat{\mathcal{F}}_B(k) > 0$, that is, buoyancy feeds energy to the system. \textcolor{black}{ Hence,  $\frac{d}{dk}\Pi_u(k) > 0$ in steady state from Eq.~(\ref{eq:energyBalance_Steady_RBC})}. \textcolor{black}{It is important to note that in stably-stratified flows, buoyancy depletes energy from the system. Thus, for such flows, $\hat{\mathcal{F}}_B < 0$, resulting in $\frac{d}{dk}\Pi_u(k) < 0$. This means that the flux decreases with wavenumber in the inertial range; this is an important ingradient of Bolgiano-Obukhov scaling~\cite{Bolgiano:JGR1959,Obukhov:DANS1959}. Since the flux does not decrease with wavenumber in thermal convection, Bolgiano-Obukhov scaling is ruled out.}

Further, in turbulent convection, \citet{Pandey:PF2016} and \citet{Pandey:PRE2016} showed that buoyancy is strong only at large scales and is weak in the inertial range. \citet{Nath:PRF2016} showed that the distribution of velocity field in turbulent convection is nearly isotropic similar to hydrodynamic turbulence, again indicating weak buoyancy.
  
Based on the above observations, \citet{Kumar:PRE2014}, and \citet{Verma:NJP2017} argued that $\hat{\mathcal{F}}_B$ does not bring about a noticeable increase in $\Pi_u(k)$ (See Fig.~\ref{fig:Flux_RBC}). \textcolor{black}{  Therefore,  $\hat{\mathcal{F}}_B \approx \hat{D}_u \approx 0$, which reduces Eq.~(\ref{eq:energyBalance_Steady_RBC}) to }
\begin{equation} 
\frac{d}{dk} \Pi_u(k) \approx 0, \quad \Rightarrow \Pi_u(k) \approx \mathrm{constant}.
\label{eq:constFlux} 
\end{equation}
\textcolor{black}{Thus, it can be inferred from Eq.~(\ref{eq:constFlux}) that Kolmogorov's theory of hydrodynamic turbulence is also applicable to thermal convection.} \textcolor{black}{Integrating Eq.~(\ref{eq:energyBalance_Steady_RBC}) from $0$ to a small wavenumber $k_0$ lying in the inertial range yields
\begin{equation}
\Pi_u(k_0)-\Pi_u(0) = \int_{0}^{k_0} \hat{\mathcal{F}}_B(k)dk-\int_{0}^{k_0}\hat{D}_u(k)dk. \label{eq:Energy_integrate_RBC_0_k} \\
\end{equation}
Since $\Pi_u(0)=0$ and the dissipation rate is negligible at small wavenumbers, the above equation reduces to
\begin{equation}
\Pi_u(k_0) \approx \int_{0}^{k_0} \hat{\mathcal{F}}_B(k)dk. \label{eq:Flux_forcing_RBC} \\
\end{equation}
Since $\hat{\mathcal{F}}_B(k)$ is strong at large scales, we deduce from Eq.~(\ref{eq:Flux_forcing_RBC}) that a large part of energy is injected by buoyancy at large scales that contributes to the energy flux in the inertial range; this feature is similar to hydrodynamic turbulence.} 

There is, however, a difference between the energetics of RBC and that of 3D turbulence.
Integrating Eq.~(\ref{eq:energyBalance_Steady_RBC}) from $k_0$ to $\infty$ yields \textcolor{black}{
\begin{equation}
\Pi_u(\infty)-\Pi_u(k_0) = \int_{k_0}^{\infty} \hat{\mathcal{F}}_B(k)dk-\int_{k_0}^{\infty}\hat{D}_u(k)dk. \label{eq:Energy_integrate_RBC_0} \\
\end{equation}
Since $\Pi_u(\infty)=0$, the above equation becomes
\begin{equation}
\Pi_u(k_0) = \int_{k_0}^{\infty}\hat{D}_u(k)dk - \int_{k_0}^{\infty} \hat{\mathcal{F}}_B(k)dk. \label{eq:Energy_integrate_RBC} \\
\end{equation}
Since $k_0$ is small compared to the dissipation range wavenumbers, we can write 
\begin{equation}
\int_{k_0}^{\infty}\hat{D}_u(k)dk \approx  \int_{0}^{\infty}\hat{D}_u(k)dk = \epsilon_u. \nonumber 
\end{equation}
\textcolor{black}{Now,$\int_{k_0}^{\infty} \hat{\mathcal{F}}_B(k)dk$ is the energy injected by buoyancy at small scales.} It must be noted that $\int_{k_0}^{\infty} \hat{\mathcal{F}}_B(k)dk > 0$ in RBC, because  $\hat{\mathcal{F}}_B(k)$, albeit weak, is positive and adds up to a significant amount when integrated over the inertial and dissipation range~(see Sec.~\ref{sec:Comparison}).} Therefore,
\begin{equation} 
\Pi_u(k_0) \approx \epsilon_u - \int_{k_0}^{\infty} \hat{\mathcal{F}}_B(k)dk < \epsilon_u. 
\label{eq:fluxRBC} 
\end{equation}
Eq.~(\ref{eq:fluxRBC}) clearly shows that unlike in hydrodynamic turbulence, the energy flux in the inertial range is smaller than the dissipation rate \textcolor{black}{due to the energy injected by buoyancy at small scales. Recall that in hydrodynamic turbulence, no energy is injected in these regimes.} In Sec.~\ref{sec:Comparison}, using the results of numerical simulations of turbulent convection, we show that the energy flux is smaller than the dissipation rate by a factor of two to three \textcolor{black}{for our selected cases. Note that this factor likely depends on Ra, Pr, type of boundary conditions, etc. A careful study of the spectra and fluxes of thermal convection for different regimes of Ra and Pr needs to be carried out to ascertain how this factor depends on the aforementioned parameters.} 

In the next subsection, following the procedure of \citet{Kolmogorov:DANS1941Dissipation,Kolmogorov:DANS1941Structure}, we derive the relation for the third-order velocity structure functions of turbulent convection.

\subsection{Velocity structure functions of turbulent convection} \label{subsec:StructureFunctions}
\textcolor{black}{ \citet{Sun:PRL2006} and \citet{Zhou:JFM2008} performed experiments of turbulent thermal convection and observed isotropy in regions away from walls.}
Using detailed numerical simulations, \citet{Nath:PRF2016}  computed the modal energy of the inertial-range Fourier modes of turbulent convection as a function of polar angle $\Theta$ (angle between buoyancy direction and the wavenumber), and found it to be approximately independent of $\Theta$.  Thus, they showed that turbulent convection is nearly isotropic. \textcolor{black}{
In Sec.~\ref{sec:SFu_main}, we compute the second-order velocity structure functions as functions of $l$ and $\Theta$ ($\Theta$ is the angle between the buoyancy direction and $l$) using our numerical data, and show that they are nearly independent of $\Theta$. This again shows near-isotropy in thermal convection. We believe that isotropy is related to the fact that in turbulent convection, buoyancy ``effectively" injects energy at large scales, but it is weak in the inertial range.} 

Further, at high Rayleigh numbers, the boundary layers are very thin, \textcolor{black}{with the boundary layer thickness $\delta_u \ll d$, $d$ being the domain height.} Therefore, for simplification, we neglect the effects of boundary layers and consider the system to be homogeneous.  \textcolor{black}{In Appendix~\ref{sec:SF_planes} we show that in turbulent thermal convection, the planar structure functions and those computed in the entire domain exhibit somewhat similar scaling;  this result too validates the assumptions of approximate homogeneity and isotropy for turbulent convection. Using the assumptions of homogeneity, isotropy and steady state,} and following similar lines of arguments as \citet{Kolmogorov:DANS1941Dissipation,Kolmogorov:DANS1941Structure}, we derive  the relation for third-order structure function for turbulent convection in the following discussion.   
 
For homogeneous and incompressible turbulent flows, the temporal evolution of the second-order velocity correlation function can be written as follows~\citealp{Kolmogorov:DANS1941Dissipation,Kolmogorov:DANS1941Structure,Frisch:book}:
\be
\frac{\partial}{\partial t}\left[\frac{1}{2}\langle u_i(\mathbf{r}) u_i(\mathbf{r+l}) \rangle \right] = T_u(\mathbf{l}) + \mathcal{F}_B(\mathbf{l}) - D_u(\mathbf{l}),
\label{eq:2orderCor_1}
\ee
where
\bea
T_u(\mathbf{l}) &=& \frac{1}{4} \nabla_l \cdot \left \langle [u(\mathbf{r+l}) - u(\mathbf{r})]^2 [\mathbf{u}(\mathbf{r+l}) - \mathbf{u}(\mathbf{r}) ] \right \rangle, \nonumber \label{eq:NonLinearEnergyTransfer} \\
\mathcal{F}_B(\mathbf{l}) &=& \langle F_i(\mathbf{r})u_i(\mathbf{r+l}) \rangle, \nonumber \label{eq:ForceCorrellation}\\
D_u(\mathbf{l}) &=&  \nu \nabla'^2 \langle u_i(\mathbf{r}) u_i(\mathbf{r+l}) \rangle. \nonumber \label{eq:Dissipation_Correllation}
\eea
Here, $T_u(\mathbf{l})$ is the non-linear energy transfer at scale $\mathbf{l}$, $\mathcal{F}_B(\mathbf{l})$ is the force correlation at $\mathbf{l}$, and $D_u(\mathbf{l})$ is the corresponding dissipation rate. The symbol $\nabla'^2$ represents the Laplacian at $\mathbf{r+l}$. Under a steady state, the left-hand side of Eq.~(\ref{eq:2orderCor_1}) disappears. Further, we focus on the inertial range where $D_u(\mathbf{l}) \approx 0$ that yields
\begin{equation}
\mathcal{F}_B(\mathbf{l}) \approx -T_u(\mathbf{l}).
\label{eq:Fb=Tu_1}
\end{equation}  
Now, $\mathcal{F}_B(\mathbf{l})$ can be expanded as Fourier series as follows:
\begin{equation}
\mathcal{F}_B(\mathbf{l}) = \sum_{\mathbf{k}} \hat{\mathcal{F}}_B (\mathbf{k}) \exp(i\mathbf{k} \cdot \mathbf{l}).
\label{eq:Force_FSeries_1}
\end{equation}
Following \citet{Verma:NJP2017}, we model $\hat{\mathcal{F}}_B(\mathbf{k})$ as~\citep{Frisch:book}
\begin{equation}
\hat{\mathcal{F}}_B(\mathbf{k})=\frac{A}{2}(\delta_{\mathbf{k,k}_0} + \delta_{\mathbf{k,-k}_0}) + Bk^{-5/3}.
\label{eq:Force_k} 
\end{equation}
Substitution of Eq.~(\ref{eq:Force_k}) in Eq.~(\ref{eq:Force_FSeries_1}) yields
\begin{align}
\mathcal{F}_B(\mathbf{l}) &= A \cos(\mathbf{k}_0 \cdot \mathbf{l}) + \int Bk^{-5/3} \exp(i\mathbf{k} \cdot \mathbf{l}) d\mathbf{k} \nonumber \\
&\approx A + DBl^{2/3}.
\label{eq:Force_FSeries_2}
\end{align}
This is because $\mathbf{k}_0\cdot \mathbf{l} \approx 0$ since turbulent convection is essentially forced by large-scale plumes~\citep{Verma:NJP2017}. Here, $B$ is a small constant. Now, for an isotropic flow, $T_u(\mathbf{l}) = T_u(l)$, and is related to the third-order structure function $S_3^u(l)$ as (see \citet{Frisch:book})
\be
T_u(l) =  \frac{1}{12} \frac{1}{l^2} \frac{d}{dl}\left[\frac{1}{l} \frac{d}{dl} \{l^4 S_3^u(l)\}\right]. \label{eq:EnergyTransfer}
\ee
Combining Eqs.~(\ref{eq:Fb=Tu_1}), (\ref{eq:Force_FSeries_2}) and (\ref{eq:EnergyTransfer}), we get
\begin{equation}
-\frac{1}{12}\frac{1}{l^2} \frac{d}{dl}\left[\frac{1}{l} \frac{d}{dl} \{l^4 S_3^u(l)\}\right]= A + DBl^{2/3}.
\label{eq:Fb=Tu_2}
\end{equation}
Integrating the above expression twice, and noting that $S_3^u(0)=0$, we obtain the following relation:
\begin{equation}
S_3^u(l)=-\frac{4}{5}(Al+D'Bl^{5/3}).
\label{eq:S_3}
\end{equation}
Now, we assume that the large-scale buoyant energy feed at $k=k_0$ equals the energy flux $\Pi_u$, and that $B$ is small. Therefore, we have $A \approx \Pi_u$, and
\begin{equation}
S_3^u(l)=-\frac{4}{5} \Pi_u l.
\label{eq:S_3_Final}
\end{equation}
Thus, the scaling of the third-order structure functions of RBC is similar to those of 3D hydrodynamic turbulence, except that $\epsilon_u$ of $S_3^u(l)$ is replaced by $\Pi_u$.  Note that $\Pi_u < \epsilon_u$ for RBC. We will verify the above relation in Sec.~\ref{sec:SFu} using numerical simulations.

It is important to note that for hydrodynamic turbulence,  $\hat{\mathcal{F}}(k)$ is provided at small wavenumbers and is equal to the viscous dissipation rate $\epsilon_u$. Inverse Fourier transform of $\hat{\mathcal{F}}(k)$ results in a constant value of $\mathcal{F}(l)$ that equals $\epsilon_u$. Using the same procedure as shown above, one can derive that $S_3^u(l) = -(4/5)\epsilon_u l$. Note that in RBC, $\epsilon_u$ of the above $S_3^u(l)$ is replaced by $\Pi_u$. We also remark that our arguments are consistent with the results of \citet{Kunnen:PRE2014}, who computed the scale-by-scale energy budget in direct numerical simulations of RBC and showed that $S_3^u(l) \neq -(4/5)\epsilon_u l$ for convective turbulence.

\textcolor{black}{Finally, as mentioned previously, it must be noted that Eq.~(\ref{eq:S_3_Final}) has been derived under the assumption of homogeneity and isotropy, which may not be the case for all regimes of turbulent convection. For example, \citet{Nath:PRF2016} has shown that anistropy is stronger for large Prandtl numbers. Thus, we cannot make the assumption of isotropy in this regime.}

In the next section, we discuss  the numerical techniques involved in the computation of the structure functions.  

\section{Numerical methods} \label{sec:Numerics}

We use two sets of numerical data to compute the velocity structure functions, each set having different boundary conditions. The first set is the data of \citet{Verma:NJP2017}, who performed direct numerical simulation (DNS) of RBC on a $4096^3$ grid. \textcolor{black}{The grid corresponds to a cube of unit dimension.} The Rayleigh and Prandtl numbers were chosen as $1.1\times10^{11}$ and unity respectively. The grid corresponds to a cubical domain of unit dimension. \textcolor{black}{The simulation was performed using a pseudo-spectral code~\citep{Verma:Pramana2013tarang,Chatterjee:JPDC2018}.} Free-slip and isothermal boundary conditions were employed at the top and bottom plates, and periodic boundary conditions were employed at the side walls. For details, refer  to \citet{Verma:NJP2017}.

The second set of data is that of \citet{Kumar:RSOS2018}. \textcolor{black}{This simulation was performed using a finite volume solver~\citep{Jasak:CD2007} on a non-uniform $256^3$ grid that corresponds to a cube of unit dimension}. The Rayleigh and Prandtl numbers were chosen as $1\times10^8$ and unity respectively. No-slip boundary conditions were imposed at all the walls; such realistic boundary conditions capture the wall effects. Isothermal boundary conditions were imposed at the top and bottom plates and adiabatic boundary conditions at the side walls. \textcolor{black}{For spatial discretization schemes, time-marching method, and the validation of the code, see Refs.~\cite{Kumar:RSOS2018,Bhattacharya:PF2018,Bhattacharya:PF2019}.} We interpolate the velocity fields to a uniform $256^3$ grid.

We compute the velocity structure functions \textcolor{black}{in the entire domain} using a combination of shared (OpenMP) and distributed memory (MPI) parallelization~(see \citet{Pacheco:book:PP}). The computations involve running six nested \texttt{for} loops: the outer three loops describing the position vector $\mathbf{r}$ and the inner three loops describing $\mathbf{r+l}$.
To save computational resources, we condense our free-slip data to $512^3$ grid. Note that we are interested only in scales pertaining to the inertial range and not the dissipative scales. After the aforementioned coarsening, we are still able to resolve scales above $6\eta$ and capture the inertial range very well in addition to avoiding unnecessary computational costs. The number of MPI nodes used were equal to the number of grid points in the $x$-direction, while the number of OpenMP threads used were 32.

In the forthcoming section, we will discuss the numerical results.

\section{Numerical results} \label{sec:SFu}

In the present section, for turbulent thermal convection, we describe the scaling of the velocity structure functions, the probability distribution functions of velocity increments, and the difference between the energy flux and viscous dissipation rate.

\begin{table*}
	 \caption{For the two simulations of RBC: Rayleigh number $\mathrm{Ra}$, Nusselt number $\mathrm{Nu}$, kinematic viscosity $\nu$, viscous dissipation rate $\epsilon_u$, and Kolmogorov length scale $\eta$.}
  \begin{ruledtabular}
\def~{\hphantom{0}}
  \begin{tabular}{cccccc} 
       Case & Ra & $\mathrm{Nu}$ & $\nu$  & $\epsilon_u$ & $\eta$ \\ \hline \\
       Free-slip & $1.1 \times 10^{11}$ & $582$ & $3.02 \times 10^{-6}$ & $2.59 \times 10^{-3}$ & $3.21 \times 10^{-4}$ \\
       No-slip & $1.0 \times 10^8$ & $32.8$ & $1.00 \times 10^{-4}$ & $3.18 \times 10^{-3}$ & $4.21 \times 10^{-3}$ \\ 
  \end{tabular}
  \label{table:Eta}
  \end{ruledtabular}
\end{table*}
\subsection{Structure functions} \label{sec:SFu_main}
\begin{table*}
	\caption{For the free-slip and no-slip simulations of RBC: prefactor $\mathcal{A}$ and the scaling exponent $\zeta_q$ for the structure functions computed by fitting the relation $|S_q^u(l)| = \mathcal{A} l^{\zeta_q} $ to our data.}
	\begin{ruledtabular}
  \begin{tabular}{ccccc}
       ~ & \multicolumn{2}{c}{Free-slip simulation ($\mathrm{Ra}=1.1 \times 10^{11}$)} & \multicolumn{2}{c}{No-slip simulation ($\mathrm{Ra}=1.0 \times 10^{8}$)} \\
       \hline 
       $q$ & $\mathcal{A}$ & $\zeta_q$ & $\mathcal{A}$ & $\zeta_q$  \\ \hline \\
       2 &  $(2.8 \pm 0.1)  \times 10^{-2}$ & $0.70 \pm 0.01$ & $(2.3 \pm 0.1) \times 10^{-2}$ & $0.71 \pm 0.01$ \\
       3 &  $(9.3 \pm 0.5) \times 10^{-4}$ & $0.97 \pm 0.01$ & $(8.5 \pm 0.5) \times 10^{-4}$ & $0.98 \pm 0.02$ \\
       4 & $(2.0 \pm 0.1) \times 10^{-3}$ & $1.26 \pm 0.02$ & $(1.6 \pm 0.1) \times 10^{-3}$ & $1.25 \pm 0.02$  \\
       5 & $(1.5 \pm 0.1) \times 10^{-4}$ & $1.45 \pm 0.02$ & $(2.6 \pm 0.2) \times 10^{-4}$ & $1.60 \pm 0.04$ \\
       6 & $(1.8 \pm 0.1) \times 10^{-4}$ & $1.69 \pm 0.02$ & $(2.6 \pm 0.2) \times 10^{-4}$ & $1.76 \pm 0.03$ \\
       7 & $(2.1 \pm 0.1) \times 10^{-5}$ & $1.81 \pm 0.02$ & $(7.6 \pm 0.8) \times 10^{-5}$ & $2.01 \pm 0.05$ \\
       8 & $(2.7 \pm 0.3) \times 10^{-5}$ & $2.09 \pm 0.03$ & $(6.0 \pm 0.6) \times 10^{-5}$ & $2.16 \pm 0.05$ \\
       9 & $(3.9 \pm 0.7) \times 10^{-6}$ & $2.14 \pm 0.05$ & $(2.6 \pm 0.4) \times 10^{-5}$   & $2.33 \pm 0.07$ \\
       10 & $(3.1 \pm 0.5) \times 10^{-6}$ & $2.28 \pm 0.05$ & $(2.1 \pm 0.3) \times 10^{-5}$ & $2.51 \pm 0.07$
  \end{tabular}
  \end{ruledtabular}
\label{table:SFDetails}
\end{table*}

\textcolor{black}{Before computing the structure functions, we first numerically compute the viscous dissipation rate $\epsilon_u$ using the velocity field data of our free-slip and no-slip cases. We use the relation 
\be
\epsilon_u = \langle 2 \nu S_{ij}S_{ij} \rangle
\ee
to compute the viscous dissipation rate, where $S_{ij}$ is the strain rate tensor, and $\langle . \rangle$ represents the volume average. Further, we compute the Kolmogorov length scale $\eta$ and the Nusselt number Nu using the following relations~\cite{Frisch:book,Pope:book,Lohse:ARFM2010}:
\begin{eqnarray}
\eta &=& \left( \frac{\nu^3}{\epsilon_u} \right)^{1/4},\\
\mathrm{Nu} &=& 1+\frac{\langle u_z \theta \rangle} {\kappa \Delta d^{-1}}.
\end{eqnarray}
}In Table~\ref{table:Eta}, we list the values of Nu, $\nu$, $\epsilon_u$ and $\eta$ for both free-slip and no-slip data. Clearly, $\eta$ is larger for the no-slip case because of lower $\mathrm{Ra}$. 
\textcolor{black}{Further, we remark that the viscous boundary layers are thin for our data, with $\delta_u=0$ for the free-slip simulation and  $\delta_u \approx 2 \eta$ for the no-slip simulation~\cite{Bhattacharya:PF2018}. 
Thus, most of the flow resides in the bulk.} 

\begin{figure}[t]
	\begin{center}
		\includegraphics[keepaspectratio=true,scale= 0.25]{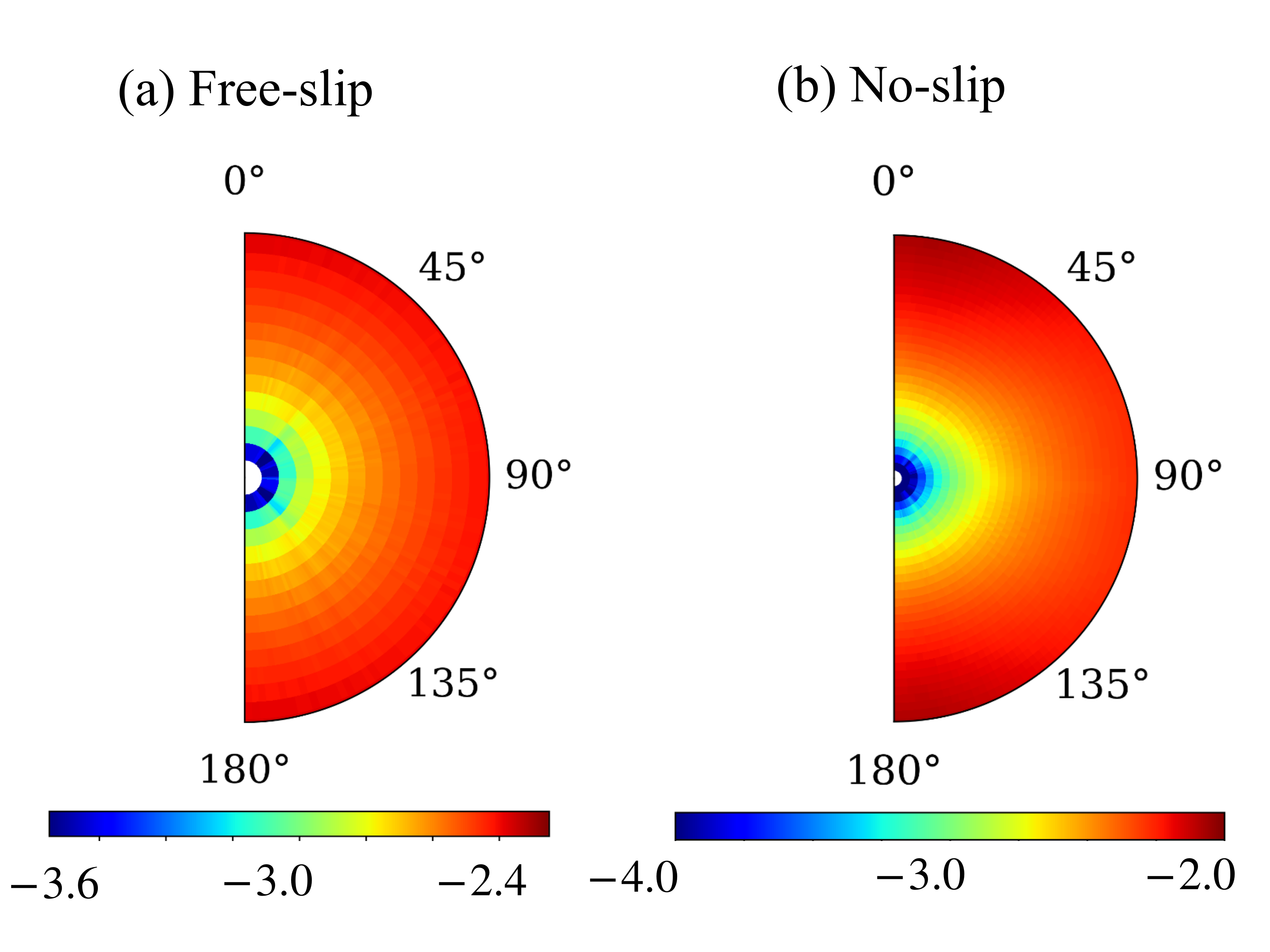}
		\caption{\textcolor{black}{For the (a) free-slip and (b) no-slip simulations of RBC: Polar ($l,\Theta$) plots of the logarithms of second-order velocity structure functions, where $\Theta$ is the angle between the buoyancy-direction and $l$. $l$ spans the inertial-dissipation range: $0 < l/\eta < 210$ for the free-slip data and $0 < l/\eta < 40$ for the no-slip data. The structure functions are nearly independent of $\Theta$, thus demonstrating near-isotropy in the inertial-dissipation range.}}
		\label{fig:thetaIsotropy}
	\end{center}
\end{figure}

\textcolor{black}{Next, we validate the assumption of isotropy in turbulent convection. 
Using both sets of data, we compute the second-order velocity structure functions  in the entire domain as functions of $l$ and $\Theta$, where $\Theta$ is the angle between the buoyancy direction and $l$.
Figs.~\ref{fig:thetaIsotropy}(a,b) exhibit the polar plots $S_2^u(l,\Theta)$, with $l$ spanning the inertial-dissipation range ($0<l/\eta<210$ for the free-slip case and $0<l/\eta<40$ for the no-slip case). 
The figures clearly show that the structure functions are nearly independent of $\Theta$, thereby demonstrating near-isotropy in the inertial-dissipation range.}

\begin{figure}[t]
\includegraphics[scale=0.395]{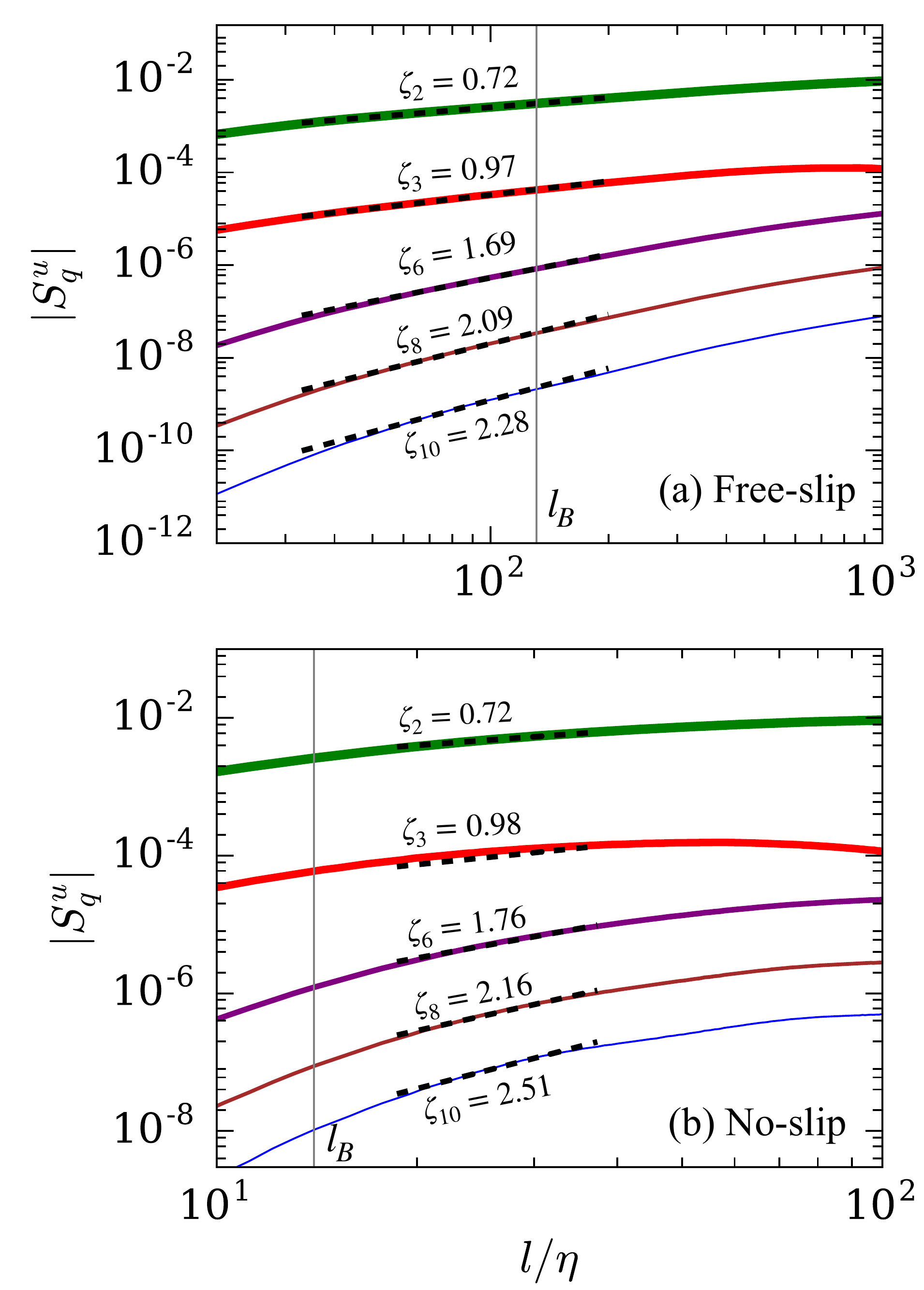} 
\caption{For (a) the free-slip and (b) no-slip simulations of RBC: plots of $|S_q^u|$ with decreasing line thickness for $q=2$ (green), $3$ (red), $6$ (purple), $8$ (brown) and $10$ (blue) vs. $l/\eta$. The vertical solid gray line marks the Bolgiano length scale.}
\label{fig:SF_U}
\end{figure}

Now, we compute the magnitude of $S_q^u$ as a function of $l$ \textcolor{black}{in the entire domain}, with $q$ ranging from 2 to 10.  Fig.~\ref{fig:SF_U} exhibits the plots of structure functions of orders 2, 3, 6, 8 and 10 versus $l/\eta$ for both sets of data.
Contrary to the results of \citet{Benzi:EPL1994,Benzi:EPL1994b}, we observe a discernible scaling range for the third order structure function. The range is found to be $32<l/\eta < 200$ for the free-slip data and $19<l/\eta < 40$ for the no-slip data.
The range is much smaller for the no-slip case because of the higher value of $\eta$. \textcolor{black}{Note that the length scales in the inertial range are much larger than the boundary layer thickness.}

\begin{figure}[htbp]
\centering
\includegraphics[scale=0.395]{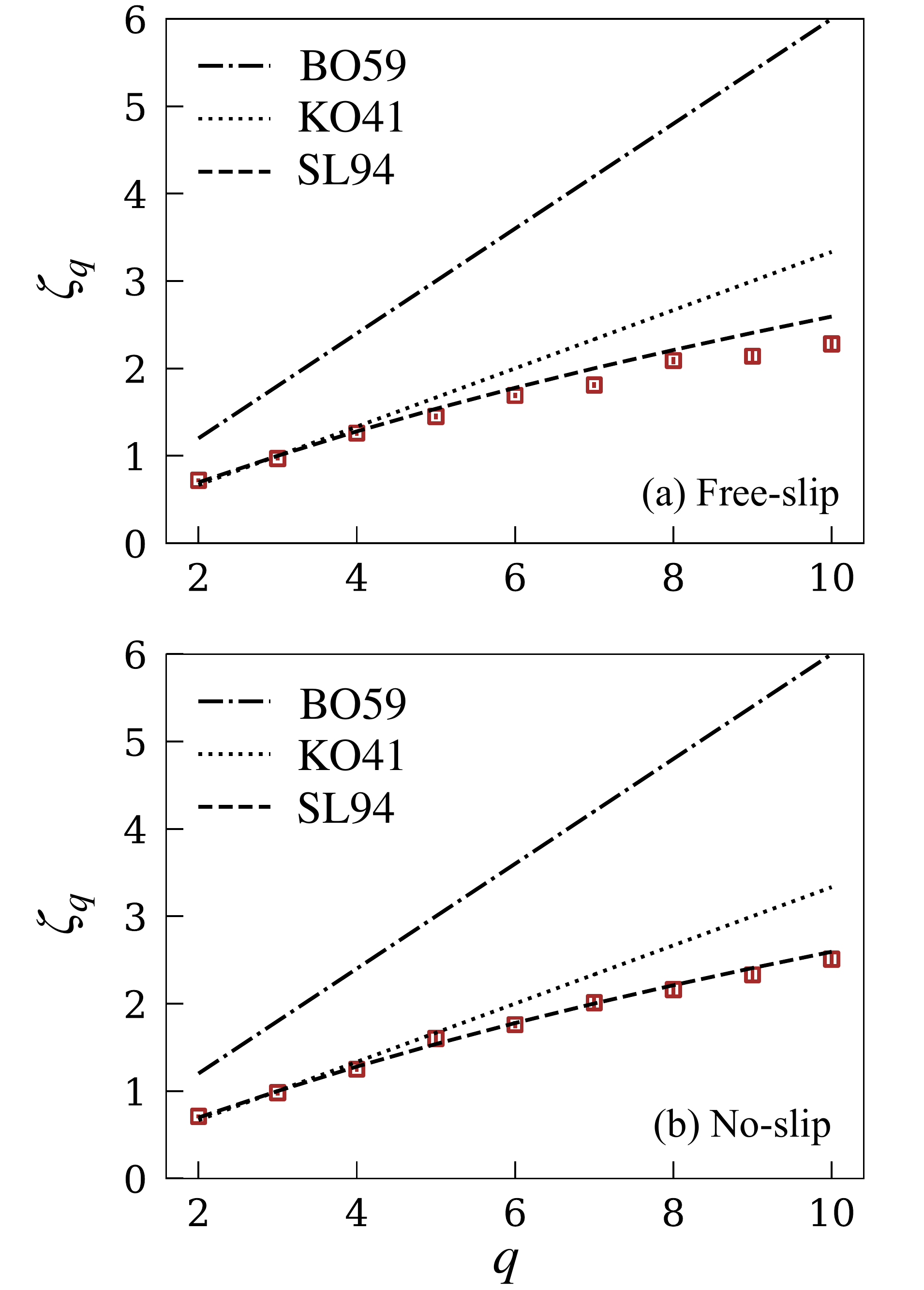}
\caption{For (a) the free-slip and (b) no-slip simulations of RBC: plots of $\zeta_q$ (squares) vs. $q$. $\zeta_q$ matches closely with the predictions of \citet{She:PRL1994}(dashed line). The figures also contain Kolmogorov's prediction $\zeta_q=q/3$ (dotted line) and Bolgiano-Obukhov's prediction $\zeta_q=3q/5$ (chained line).}
\label{fig:SheLeveque}
\end{figure}

We compute the scaling exponents $\zeta_q$ and the prefactor $\mathcal{A}$ by fitting the relation $S_q^u(l) = \mathcal{A}l^{\zeta_q}$ to our data within the scaling range. Table \ref{table:SFDetails} lists $\mathcal{A}$ and $\zeta_q$ for both sets of data. Note that $\zeta_3=0.97$ and $0.98$ for the free-slip and the no-slip cases respectively, which are close to Kolmogorov scaling of $S_3^u \sim l$.
From Table~\ref{table:SFDetails} and Figs.~\ref{fig:SF_U} and \ref{fig:SheLeveque}, we observe that for lower orders, the scaling exponents $\zeta_q$  for  free-slip and no-slip boundary conditions are nearly equal, and they are close to $q/3$, which is a generalisation of Kolmogorov's theory of turbulence. For $q=2$, $\zeta_2 \approx 2q/3$ that yields $k^{-5/3}$ energy spectrum.  These results are consistent with the Kolmogorov energy spectrum in thermal convection observed by \citet{Kumar:PRE2014}, \citet{Verma:NJP2017}, and \citet{Kumar:RSOS2018}. 
\begin{figure}
\centering
\includegraphics[scale=0.4]{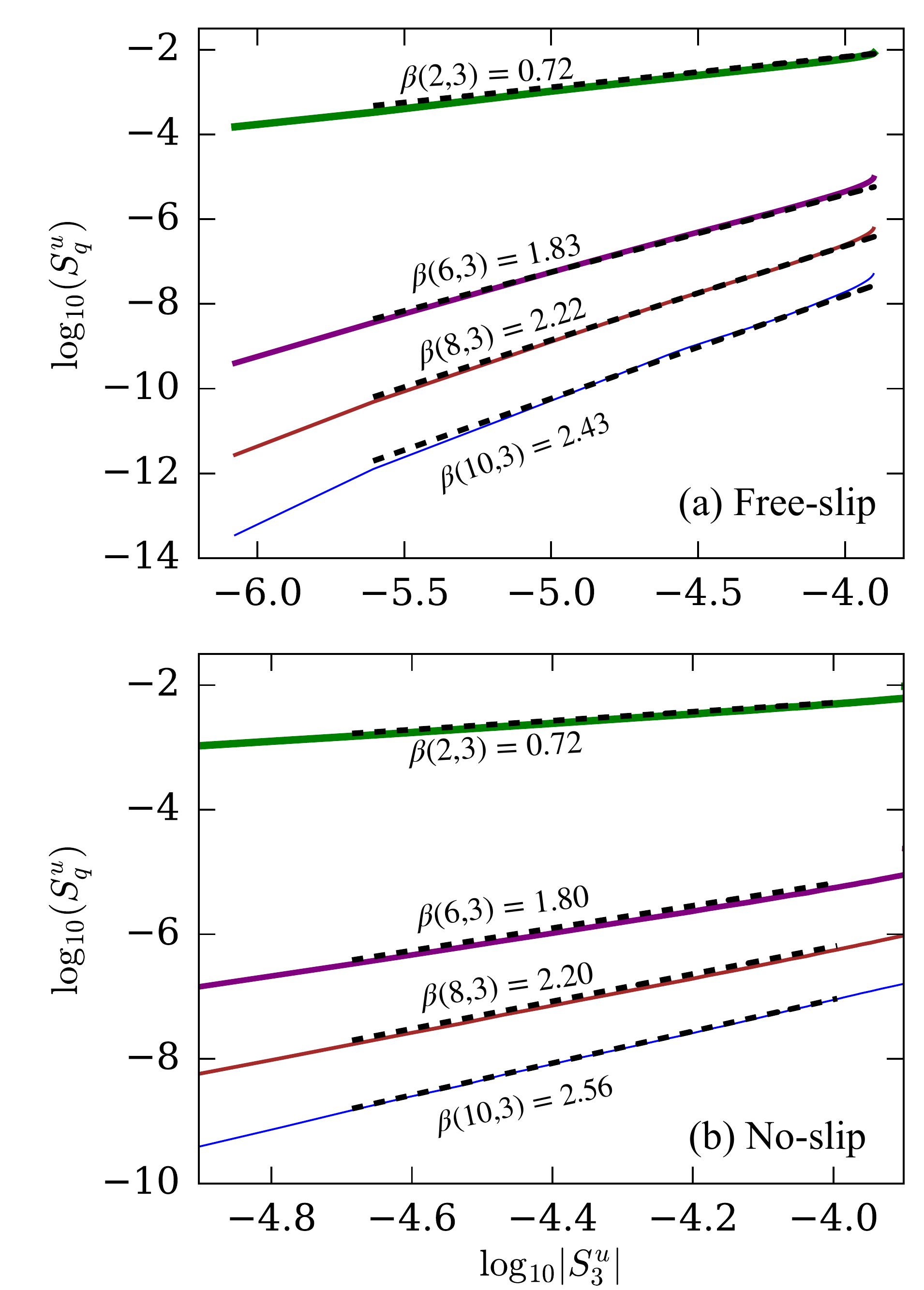}
\caption{For (a) the free-slip and (b) no-slip simulations of RBC: plots of $S_q^u$ vs. $S_3^u$. This extended self-similarity goes beyond the inertial range.}
\label{fig:SF_U_ESS}
\end{figure} 
Our results are also consistent with those of \citet{Sun:PRL2006} and \citet{Kaczorowski:JFM2013}, who report Kolmogorov scaling of the structure functions of RBC computed at the cell center. 
On the other hand, our results are contrary to those of \citet{Benzi:EPL1994,Benzi:EPL1994b},  \textcolor{black}{\citet{Calzavarini:PRE2002}}, and \citet{Kunnen:PRE2008}~(for $\mathrm{Ra=10^8}$), who deduce Bolgiano-Obukhov scaling based on their simulations. 
However, it must be noted that \citet{Kunnen:PRE2008} could not observe Bolgiano-Obukhov scaling for $\mathrm{Ra>10^8}$; rather, they report Kolmogorov scaling, similar to our results. We will discuss more on Bolgiano-Obukhov scaling later in this section.   

As illustrated in Table~\ref{table:SFDetails} and Fig.~\ref{fig:SheLeveque},  higher order $\zeta_q$'s for the free-slip data are marginally lower than those for the no-slip data.   Also, for higher order structure functions, $\zeta_q$ deviates from $q/3$ due to intermittency.   To explain intermittency effects in hydrodynamic turbulence, \citet{She:PRL1994} proposed the following   model for $\zeta_q$:
\begin{equation}
\zeta_q=\frac{q}{9}+2\left(1-\left(\frac{2}{3} \right)^{q/3} \right). \label{eq:She-Leveque}
\end{equation} 
Interestingly, the aforementioned equation describes $\zeta_q$ calculated using our RBC data quite well; see Figs.~\ref{fig:SheLeveque}(a,b). These results demonstrate similarities between  $\zeta_q$ scaling in convection and in hydrodynamic turbulence,  consistent with earlier results  \citep{Kumar:PRE2014, Verma:NJP2017, Verma:book:BDF}. Our results also match with the experimental work of \citet{Sun:PRL2006}, who observed the scaling exponents of structure functions calculated at cell-center to fit with She-Leveque's model. 

In Fig.~\ref{fig:SF_U_ESS}, we plot the logarithms of $S_2^u$, $S_6^u$, $S_8^u$ and $S_{10}^u$ versus $\log_{10}|S_3^u|$ for both free-slip and no-slip cases, and observe the structure functions to be self-similar, that is,
\begin{equation}
S_q^u \sim (S_3^u)^{\beta(q,3)}, \label{eq:ESS}
\end{equation}
where $\beta(q,3)=\zeta_q/\zeta_3$. The computed values of the exponent $\beta(q,3)$ are also shown in the figure. This scaling occurs for $l/\eta$ ranging from $12$ to $530$ for the free-slip case and $9$ to $45$ for the no-slip case. The range of $S_q^u$ versus $S_3^u$ plots of Fig.~\ref{fig:SF_U_ESS} is wider than that of $S_q^u$ plots of Fig.~\ref{fig:SF_U} (In Fig.~\ref{fig:SF_U_ESS}, the range extends well beyond the inertial range to the dissipative scales). This is called extended self-similarity (ESS)~\cite{Benzi:PRE1993,Chakraborty:JFM2010}. ESS has been observed in previous studies of convection~\cite{Benzi:EPL1994,Benzi:EPL1994b,Lohse:ARFM2010}.  Note that ESS was first reported by \citet{Benzi:PRE1993} in hydrodynamic turbulence. 

According to \citet{Pope:book}, the upper limit of the inertial range can be estimated by $l_{EI}^P \approx L/6$ and the lower limit $l_{DI}^P \approx 60 \eta$. Going by this estimate, $l_{EI}^P = 530\eta$ for our free-slip data. Note that the upper and the lower limits of the power-law range of the structure functions for our free-slip data are of the same order of magnitude as Pope's estimate. For the no-slip case, because of the large value of $\eta$ and the dissipative nature of OpenFOAM solver, $l_{DI}^P(=60 \eta)$ is greater than $l_{EI}^P(=40\eta)$. Therefore, Pope's estimate for the lower limit does not hold for the no-slip case; this is expected because Pope's estimates are expected to work for homogenous and isotropic turbulence, or periodic boundary condition.

An important point to note is that  $\zeta_q$ curve does not fit with $\zeta_q=3q/5$, which is a generalisation of   Bolgiano-Obukhov (BO) model.   As discussed in Sec.~\ref{subsubsec:ConvectionPhenomenology}, \citet{Kumar:PRE2014}, and \citet{Verma:NJP2017} have argued against Bolgiano-Obukhov (BO) model for RBC based on energy flux arguments.  This result is contrary to some of the earlier works~\cite{Kunnen:PRE2008,Benzi:EPL1994,Benzi:EPL1994b,Ching:PRE2000,Procaccia:PRL1989,Lvov:PRL1991,Lvov:PD1992} that argue in favour of  Bolgiano-Obukhov model.   Note that Bolgiano length computed using $l_B = \mathrm{Nu^{1/2}/(PrRa)^{1/4}}$ are approximately $130\eta$ and $14\eta$ for the free-slip and no-slip boundary conditions respectively.  They are marked as vertical  lines in Fig.~\ref{fig:SF_U}.  We do not discuss $l_B$ in detail because Bolgiano-Obukhov (BO) model has been shown to be inapplicable for RBC~\citep{Kumar:PRE2014, Verma:NJP2017, Verma:book:BDF}\textcolor{black}{(see Sec.~\ref{subsubsec:ConvectionPhenomenology})}.
 
   
\textcolor{black}{In Appendix~\ref{sec:SF_planes},  we compute the planar structure functions for several horizontal cross sections.  We observe that the these structure functions are somewhat similar to those  described above, with a  difference that planar structure functions exhibit relatively higher fluctuations.  This is due to lesser averaging for the planar structure function.}
    
In the next subsection we describe the probability distribution function (PDF) for the velocity difference between two points.

\subsection{Probability distribution function for velocity increments}
\begin{figure}[b]
\centering
\includegraphics[scale=0.4]{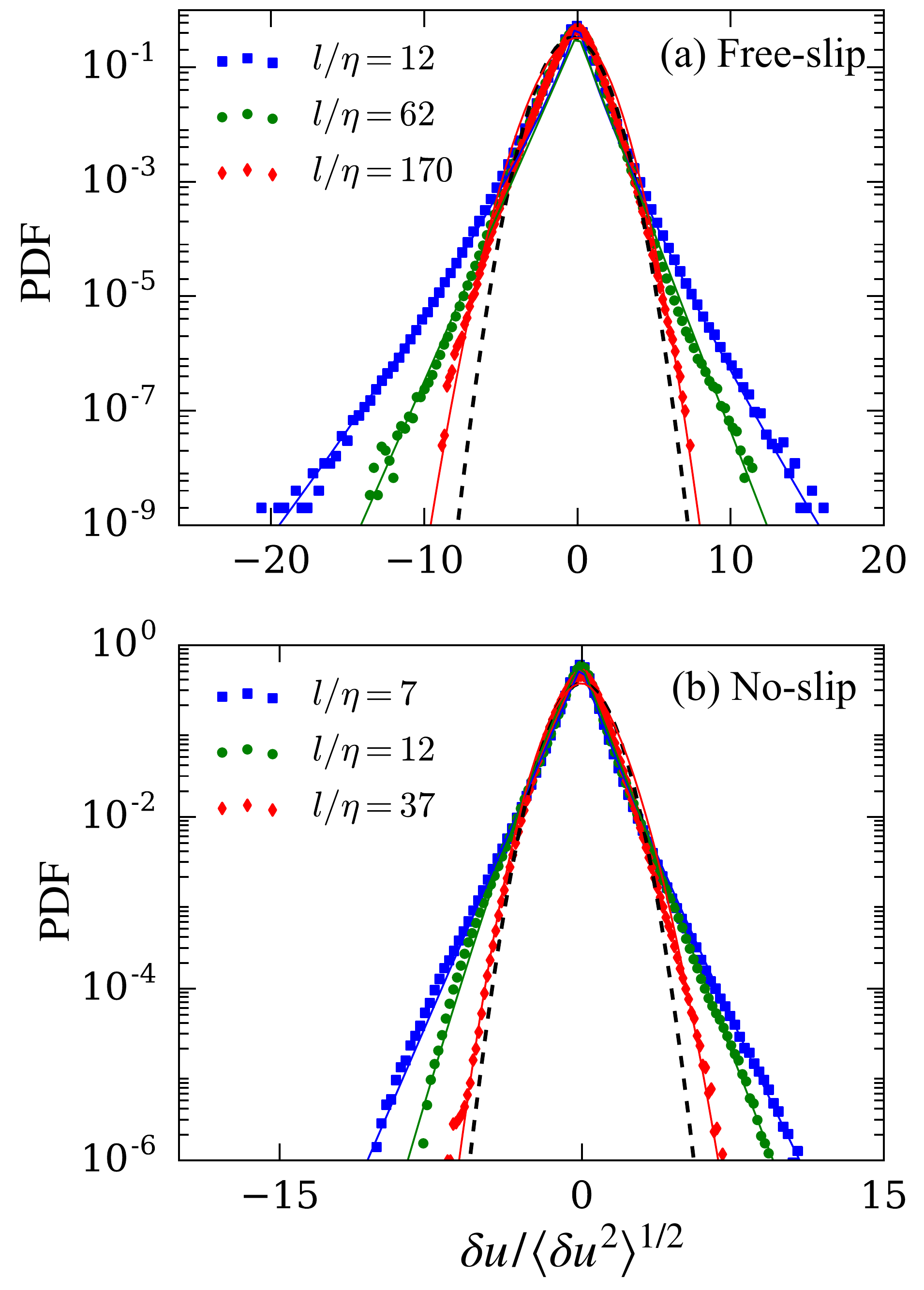}
\caption{For (a) the free-slip and (b) no-slip simulations of RBC: probability distribution functions of $\delta u$ for various $l/\eta$ (as shown in legends). The tails fit well with stretched exponential (solid curves). The dashed black curves represent the standard Gaussian distribution.}\label{fig:PDFs}
\end{figure}
For different values of $l/\eta$, we compute the probability distribution functions (PDFs) of velocity increments, $\delta u=\{\mathbf{u(r+l)-u(r)}\} \cdot \hat{\mathbf{l}}$, using the free-slip and the no-slip data. 
Fig.~\ref{fig:PDFs}(a) exhibits the PDFs of $\delta u$ for the free-slip data. For small $l$, the PDFs are non-Gaussian with wide tails. The tails fit with a stretched exponential curve given by $P(\delta u) \sim \exp(-m |\delta u^*|^\alpha)$, where $\delta u^*=\delta u / \sqrt{\langle \delta u^2 \rangle}$. We observe that the stretching exponent $\alpha=0.8$, $1.0$, and $1.8$ for $l/\eta=12$, $62$, and $170$ respectively. Thus, the PDFs become closer to Gaussian (represented by dashed black curve) as $l$ increases. This is expected since the velocities at two largely separated points become independent of each other. Our results are similar to those observed in hydrodynamic turbulence~(see Refs~\cite{Kailasnath:PRL1992,Donzis:PF2008}).

Fig.~\ref{fig:PDFs}(b) exhibits the PDFs of $\delta u$ calculated using the no-slip data. Clearly, the tails are narrower compared to the free-slip case. This is because of the weaker velocity fluctuations owing to the lower Rayleigh number. Moreover, the presence of viscous boundary layers also reduces the fluctuations.  \citet{Pandey:Pramana2016} show that for the same parameters, the large scale velocity and heat flux are less for convection with no-slip walls than with free-slip walls. Similar to the free-slip case, the tails of the PDFs fit well with a stretched exponential. For $l/\eta=7$, $12$, and $37$, $\alpha$'s are $0.9$, $1.0$, and $1.7$ respectively for the left tail, and $1.0$, $1.2$, and $1.9$ respectively for the right tail. The PDFs become close to Gaussian at large scales, similar to the free-slip case.

\subsection{Buoyancy forcing, energy flux and viscous dissipation rate} \label{sec:Comparison}
\begin{table*}
	\caption{For the two simulations of RBC: energy flux $\Pi_u$ computed using the third-order structure functions, viscous dissipation rate $\epsilon_u$, and the Kolmogorov constant $K_{\mathrm{KO}}$.}
  \begin{ruledtabular}
\def~{\hphantom{0}}
  \begin{tabular}{cccc}

       Case & $\Pi_u$  & $\epsilon_u$ & $K_{\mathrm{KO}}$ \\
       \hline \\
        Free-slip & $(1.29 \pm 0.06) \times 10^{-3}$ & $2.59 \times 10^{-3}$ & $1.59 \pm 0.09$ \\
        No-slip & $(1.09 \pm 0.03) \times 10^{-3}$ & $3.18 \times 10^{-3}$ & $1.53 \pm 0.04$ \\
     
  \end{tabular}
  \label{table:Energetics}
  \end{ruledtabular}
\end{table*}
In this section, we provide a numerical demonstration that  the energy flux and the viscous dissipation rate differ in RBC.

Using the third-order velocity structure functions, we calculate the energy flux $\Pi_u$ using Eq.~(\ref{eq:S_3_Final}) as
\begin{equation}
\Pi_u = -\frac{5}{4}\frac{S_3^u}{l}.
\label{eq:FluxFromS3}
\end{equation} 
We list the values of the energy flux in Table~\ref{table:Energetics}. \textcolor{black}{We also compute the Fourier transform of our velocity and temperature field data, and compute the spectral energy flux using the following relation~\cite{Verma:Pramana2013tarang,Chatterjee:JPDC2018}:}
\begin{equation} \textcolor{black}{
\Pi_u(k_0) = \sum_{k \geq k_0} \sum_{p<k_0} \delta_\mathbf{k,p+q} \Im(\mathbf{[k \cdot u(q)][u^*(k) \cdot u(p)]}).}
\label{eq:flux_MtoM}
\end{equation}

\begin{figure}[b]
\centering
\includegraphics[scale=0.43]{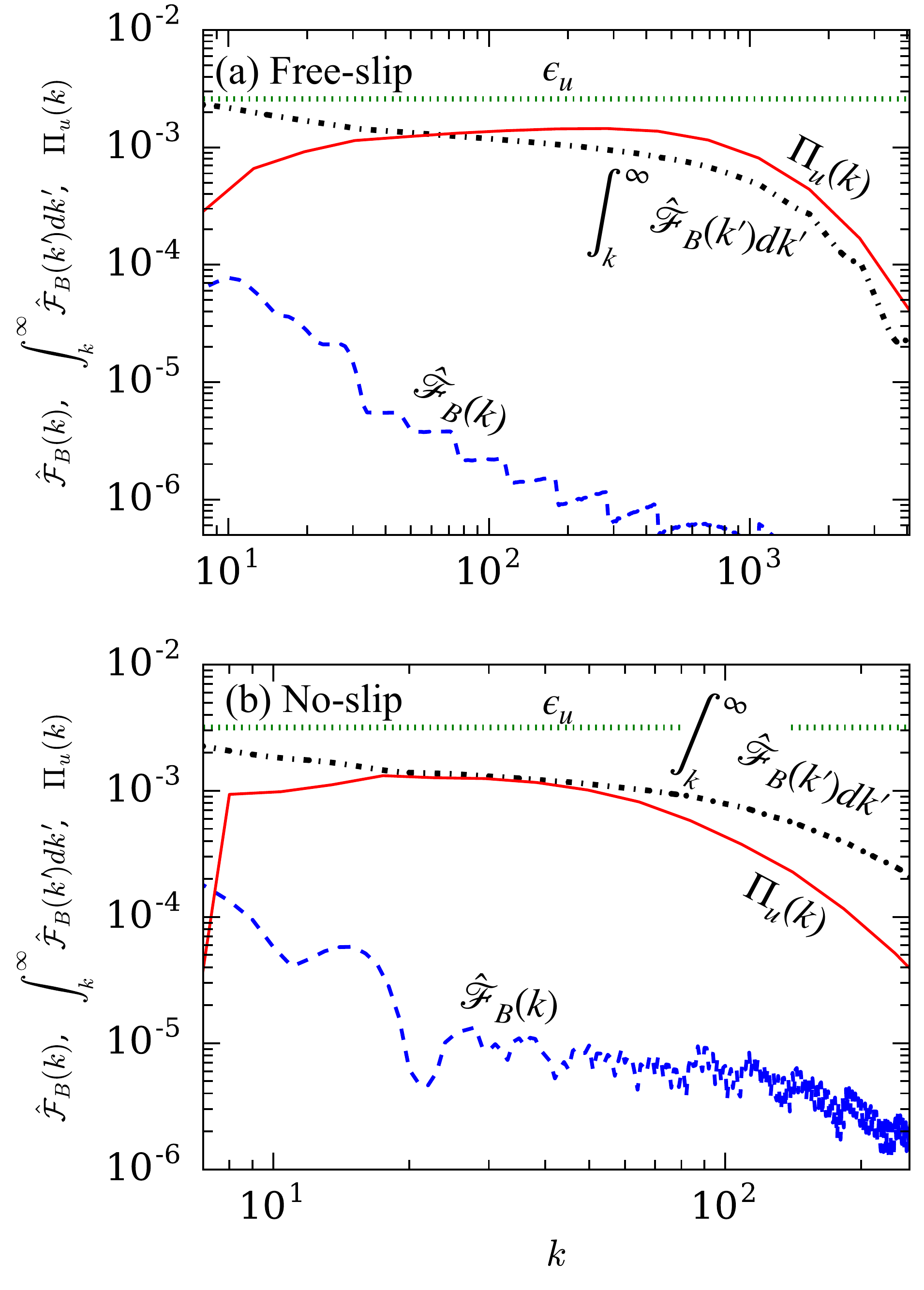}
\caption{\textcolor{black}{For (a) the free-slip and (b) no-slip simulations of RBC: the spectra of buoyancy forcing $\hat{\mathcal{F}}_B(k)$ (dashed blue lines), its integral $\int_k^\infty \hat{\mathcal{F}}_B(k')dk'$ (chained black lines), and the kinetic energy flux $\Pi_u(k)$ (solid red lines). $\hat{\mathcal{F}}_B(k)$ is weak in the inertial range. $\Pi_u(k)$ is of the same order as $\int_k^\infty \hat{\mathcal{F}}_B(k')dk'$ and is less than the viscous dissipation rate $\epsilon_u$ (dotted green lines).}}
\label{fig:Forcing_flux}
\end{figure}

\textcolor{black}{We plot the flux [computed using Eq.~(\ref{eq:flux_MtoM})] against $k$ in Fig.~\ref{fig:Forcing_flux}. We observe the value of the flux to be almost constant in the inertial range and it closely matches with that computed using Eq.~(\ref{eq:FluxFromS3}). } In Table~\ref{table:Energetics}, we also list the values of $\epsilon_u$ computed in Sec.\ref{sec:SFu_main}. 

From the table, we observe that $\epsilon_u \approx 2 \Pi_u$ for the free-slip case and $ \approx 3 \Pi_u$ for the no-slip case. This is unlike in 3D hydrodynamic turbulence  in which flux and viscous dissipation rate are equal. Our results are consistent with our arguments in Sec.~\ref{subsubsec:ConvectionPhenomenology} where we show that the difference between the flux and the viscous dissipation rate arises due to non-zero buoyancy in the inertial range. 

\textcolor{black}{Using the values of $\Pi_u(k)$ computed using Eq.~(\ref{eq:flux_MtoM}), we numerically
compute $\frac{d}{dk} \Pi_u(k)$ using central-difference method. We also
compute the energy spectrum $E_u(k)$ and obtain the spectrum of viscous
dissipation using the relation $\hat{D}_u(k) = 2\nu k^2 E_u(k)$.
Using the values of the dissipation spectrum and $\frac{d}{dk} \Pi_u (k)$ and assuming steady state, we compute $\hat{\mathcal{F}}_B(k)$ using Eq.~(\ref{eq:energyBalance_Steady_RBC}):
\begin{equation}
\hat{\mathcal{F}}_B(k) = \frac{d }{d k} \Pi_u(k) + \hat{D}_u(k). \nonumber
\end{equation} 
}
\textcolor{black}{We plot the values of $\Pi_u(k)$, $\hat{\mathcal{F}}_B(k)$, and $\int_k^\infty \hat{\mathcal{F}}_B(k')dk'$ in Fig~\ref{fig:Forcing_flux}(a) for the free-slip case and in Fig~\ref{fig:Forcing_flux}(b) for the no-slip case. In each of the plots, we also draw a horizontal line denoting the viscous dissipation rate.}
\textcolor{black}{As shown in Figs. \ref{fig:Forcing_flux}(a,b), in the inertial range,
\be
\Pi_u \sim  \int_k^\infty \hat{\mathcal{F}}_B(k')dk', \nonumber
\ee
and is approximately  $\epsilon_u/2$ for the free-slip case and  $\epsilon_u/3$ for the no-slip case. Also, $\hat{\mathcal{F}}_B(k')$ in the inertial range is weak, consistent with our previous arguments.}
\begin{figure}[t]
	\includegraphics[scale=0.43]{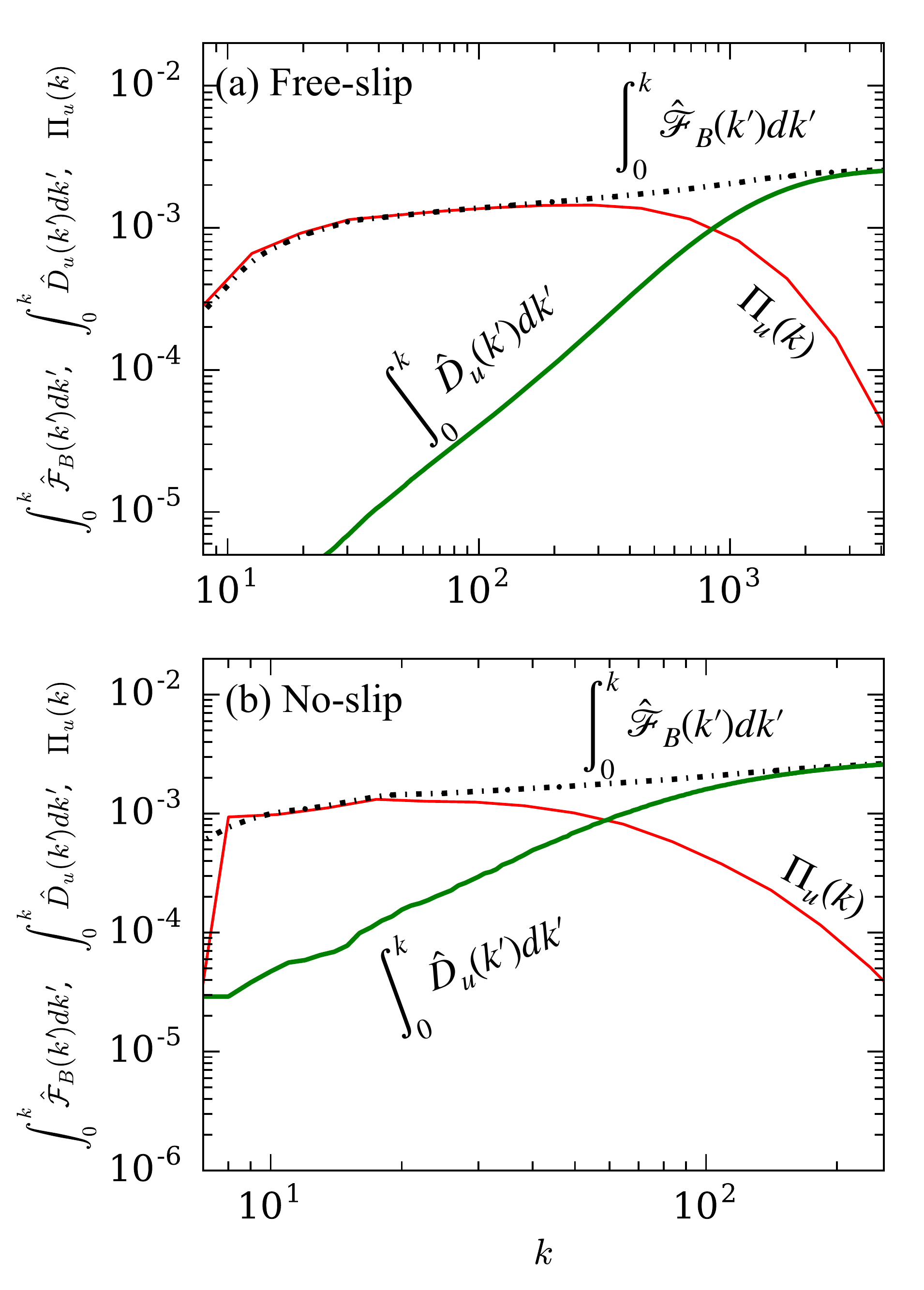}
	\caption{\textcolor{black}{For (a) the free-slip and (b) no-slip simulations of RBC: the spectra of cumulative buoyancy forcing $\int_0^k \hat{\mathcal{F}}_B(k')dk'$ (chained black lines),  kinetic energy flux $\Pi_u(k)$ (solid red lines) and cumulative dissipation rate $\int_0^k \hat{D}_u(k')dk'$ (thick green lines). The cumulative buoyancy forcing at small wavenumbers contributes mainly to the flux in the inertial range.}}
	\label{fig:Forcing_diss}
\end{figure}

\textcolor{black}{In Fig.~\ref{fig:Forcing_diss}, we plot the cumulative buoyant energy forcing $\int_0^k \hat{\mathcal{F}}_B(k')dk'$, the cumulative viscous dissipation rate $\int_0^k \hat{D}_u(k')dk'$, and the energy flux $\Pi_u(k)$ against $k$ for both sets of data. The plots clearly show that the cumulative buoyant enegy forcing at small wavenumbers contributes to the energy flux in the inertial range, consistent with our arguments in Sec.~\ref{subsubsec:ConvectionPhenomenology}. For the free-slip data, $\int_0^k \hat{\mathcal{F}}_B(k')dk'$ remains close to the flux till $k=200$, after which it deviates  from $\Pi_u(k)$. Similar behavior is also observed for the no-slip data, but with the threshold wavenumber $k=18$. Above these wavenumbers, $\int_0^k \hat{\mathcal{F}}_B(k')dk'$ increases slowly and merges with the cumulative dissipation rate $\int_0^k \hat{D}_u(k')dk'$ at dissipation wavenumbers. It is clear that $\int_0^k \hat{\mathcal{F}}_B(k')dk'$ at small wavenumbers (which contributes to the inertial range energy flux) is respectively $1/2$ and $1/3$ of the total energy injection rate ($\int_0^\infty \hat{\mathcal{F}}_B(k')dk'$) for the free-slip and the no-slip data.}  

Lastly, we compute the Kolmogorov constant $K_{\mathrm{KO}}$ by first calculating the constant $C$ using the following relation involving the second-order structure function \textcolor{black}{and the energy flux}:
\begin{equation}
S_2^u(l) = C (\Pi_u)^{2/3} l^{2/3}.
\label{eq:KoConstant}
\end{equation}
After this, we compute the Kolmogorov constant using~\citep{Pope:book}
\begin{equation}
K_{\mathrm{KO}} = \frac{55}{72}C.
\end{equation}
We list the values of Kolmogorov constant for both free-slip and no-slip cases in Table~\ref{table:Energetics}. Interestingly, $K_\mathrm{KO}$ of Table~\ref{table:Energetics} is quite close to that for hydrodynamic turbulence~\citep{Frisch:book}.

\section{Conclusions} \label{sec:Conclusions}
Using the numerical data of thermal convection, we compute the velocity structure functions $S_q^u$ for $q=2$ to $10$.  The first data set~\cite{Verma:NJP2017} was generated with free-slip boundary conditions for $\mathrm{Ra}=1.1\times 10^{11}$ and $\mathrm{Pr}=1$, and  the second set~\cite{Kumar:RSOS2018} with no-slip boundary conditions with $\mathrm{Ra}=1\times 10^8$ and $\mathrm{Pr}=1$.  We  calculate the scaling exponent $\zeta_q$ from $S_q^u$.

We show that the third-order structure functions, computed using both sets of data, scale according to Kolmogorov's theory [$S_3^u = -(4/5)\Pi_u l$]. Our results are consistent with Kolmogorov's energy spectrum observed in turbulent convection. The exponents of the structure functions of thermal convection match well with She-Leveque's predictions. We demonstrate that the structure functions show extended self-similarity. 

We also calculate the probability distribution function (PDF) of velocity increments for different values of the separation distance $l$. We show that for small $l$, the PDFs are non-Gaussian with wide tails. With increasing $l$, the PDFs become closer to Gaussian. The tails of the PDFs follow a stretched exponential, and the stretching exponent increases with $l$. Note that the PDFs of hydrodynamic turbulence show similar behaviour.

We compute the energy flux $\Pi_u$ using the third-order structure functions and show that $\Pi_u \ne \epsilon_u$; instead, it is two to three times less than $\epsilon_u$ \textcolor{black}{for our cases}. This is unlike in hydrodynamic turbulence where flux equals the dissipation rate. Using phenomenological arguments, we have shown that this difference arises due to non-zero, albeit weak, buoyancy present in the inertial range. 

In summary, the scaling behaviour of velocity structure functions of turbulent convection shows similarities with those of 3D hydrodynamic turbulence. We do not analyze the temperature structure functions in this paper. Some of the notable works on temperature structure functions of turbulent convection include those of \citet{Ching:PRE2000} and \citet{Ching:PRE2013}. We will discuss the scaling of temperature structure functions in a future work.

\section*{Acknowledgments}
We are grateful to A. Kumar and A. Chatterjee for sharing their numerical data with us. We acknowledge R. Samuel and M. Sharma for their contributions in the development of the code to calculate structure functions. We thank \textcolor{black}{S. Chakraborty and} S. Vashishtha for useful discussions. Our numerical simulations were performed on Shaheen II at {\sc Kaust} supercomputing laboratory, Saudi Arabia, under the project k1052.  This work was supported by the research grants PLANEX/PHY/2015239 from Indian Space Research Organisation, India, and by the Department of Science and Technology, India (INT/RUS/RSF/P-03)  for the Indo-Russian project.

\appendix
\section{\textcolor{black}{Extent of homogeneity in turbulent convection}} \label{sec:SF_planes}
\begin{figure}[t]
\begin{center}
\includegraphics[keepaspectratio=true,scale= 0.4]{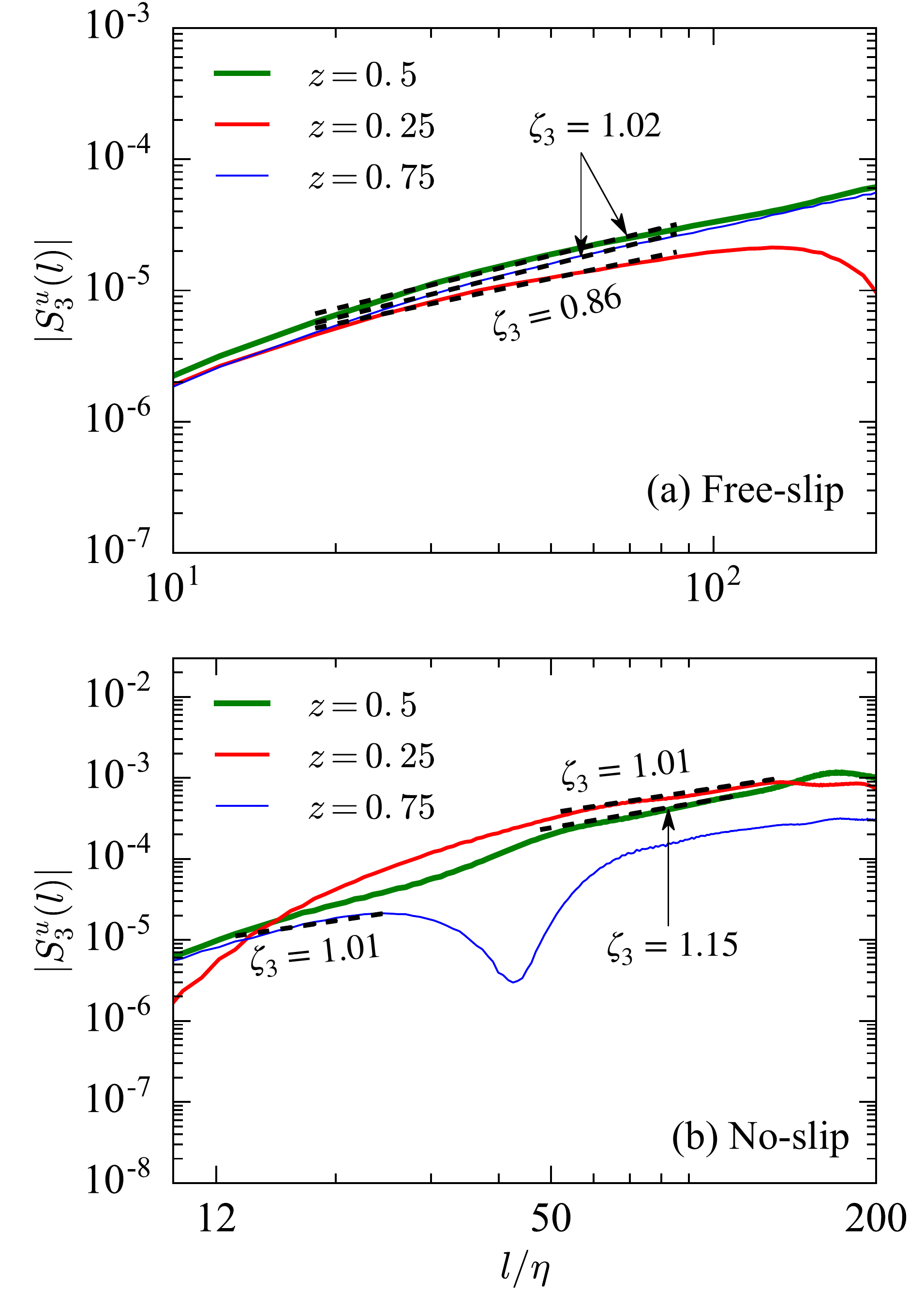}
\caption{Plots of the planar structure function $S_3^u(l)$ at $z=0.25, 0.5, 0.75$ for (a) the free-slip and (b) no-slip simulations. Despite some spatial inhomogeneity, especially for the no-slip boundary condition (b), there exist scaling range with the scaling exponent $\zeta_3 \sim 1$.}
\label{fig:SFPlan}
\end{center}
\end{figure}
For very high Rayleigh number RBC, the boundary layers are quite thin.  Hence, the flow, mostly residing in the bulk, is nearly homogeneous. However, for relatively smaller Ra (around $10^8$), there can be some inhomogeneity due to plumes and large-scale structures. To test the extent of inhomogeneity, we compute the third-order velocity structure functions for three horizontal slices of the free-slip and no-slip flow profiles detailed in the main text.  The three slices are at $z=0.25, 0.5, 0.75$.  Note that the $z=0.5$ corresponds to the mid plane.

Figure~\ref{fig:SFPlan} exhibits the plots of $|S_3^u(l)|$ vs. $l/\eta$ for the three planes. For the free-slip data with higher Ra~[Fig.~\ref{fig:SFPlan}(a)], $|S_3^u(l)| \sim l^{\zeta_3}$, where $\zeta_3 \approx 1$ for $z=0.25$ and $0.5$. However, for the $z=0.75$ plane, $\zeta_3=0.86$, which is slightly below unity. The upper and the lower limits of the scaling range are nearly same as those for the structure functions computed in the entire domain~[see Fig.~\ref{fig:SF_U}(a)]. Thus, it is reasonable to assume the free-slip data to be homogeneous.  However we observe stronger spatial inhomogeneities for the no-slip data, as shown in Fig.~\ref{fig:SFPlan}(b).  Here, the scaling regime of $|S_3^u(l)| \sim l^{\zeta_3}$  is observed for all the three planes, with $\zeta_3$ ranging from 1.02 to 1.15. However, the range of the scaling regime differs for the three planes.  Note that the spatial inhomogeneities are stronger for no-slip boundary condition due to the relatively stronger  plumes for the no-slip boundaries.

As mentioned earlier, the observed inhomogeneity, which is more prominent for no-slip data, can be attributed to localized plumes. Thus, the structure functions are required to be averaged over more points to cancel out the effects of the plumes.  That is why bulk structure functions are smoother than those for the planes, and they are closer to the predictions of She-Leveque~\cite{She:PRL1994}.

%

\end{document}